\documentstyle[preprint,aps,pre,epsfig,amsmath]{revtex}
\begin{document}
\title{Phase Transitions of Hard Disks in External Periodic
Potentials: A Monte Carlo Study}

\author{W. Strepp$^1$, S. Sengupta$^2$, P. Nielaba$^1$}

%\affiliation{$^1$ Physics Department, University of Konstanz, 
\address{$^1$ Physics Department, University of Konstanz, 
Fach M 691, 78457 Konstanz, Germany \\
$^2$ S.N. Bose National Centre for Basic Sciences,
Block JD, Sector III, Salt Lake, Calcutta 700098, India}                       

\maketitle

\date{\today}

\begin{abstract}
The nature of freezing and melting transitions for a system of
hard disks in a spatially periodic external potential is studied
using extensive Monte Carlo simulations. Detailed finite size 
scaling analysis of various thermodynamic quantities like the 
order parameter, its cumulants etc. are used to map the phase 
diagram of the system for various values of the density and 
the amplitude of the external potential. We find clear indication
of a re~-entrant liquid phase over a significant region of the parameter space.
Our simulations
therefore show that the system of hard disks behaves in a fashion similar
to charge stabilized colloids which are known to undergo an initial 
freezing, followed by a re~-melting transition 
as the amplitude of the imposed, modulating 
field produced by crossed laser beams is steadily increased.
Detailed analysis of our data shows several features consistent with a
recent dislocation unbinding theory of laser induced melting.                                                     

\end{abstract}

%\maketitle

%\begin{multicols}{2}

\newpage
\section{Introduction}
The liquid~-solid transition in systems of particles under the influence of
external modulating potentials has recently attracted a fair amount of 
attention from experiments~\cite{CAK,CAC,LA,bech,BWL,BBL,maret}, 
theory~\cite{CKS,FNR}
and computer simulations~\cite{CKSS,DK,DSK,DSK2}. This is partly 
due to the fact that well controlled, clean experiments can be performed 
using colloidal particles~\cite{col} confined between glass plates (producing
essentially a two~-dimensional system) and subjected to a spatially periodic 
electromagnetic field generated by interfering two, or more, crossed 
laser beams.  One of the more surprising results of these studies, where a  
commensurate, one dimensional, modulating potential is imposed, 
is the fact that there
exist regions in the phase diagram over which one observes 
re~-entrant~\cite{bech,BWL,BBL} 
freezing/melting behaviour. As a function of the laser field intensity the
system first freezes from a modulated liquid to a two dimensional triangular 
solid -- a further increase of the intensity confines the particles strongly 
within the troughs of the external potential, making the system 
quasi~-one~-dimensional 
which increases fluctuations and leads to re~-melting.

Our present understanding of this curious phenomenon has come from early 
mean~-field density functional~\cite{CKS} and more recent dislocation
unbinding~\cite{FNR} calculations. The mean field theories neglect 
fluctuations and therefore cannot explain re~-entrant behaviour. The order
of the transition is predicted to be first order for small laser field 
intensities, though for certain combinations of external potentials (which 
includes the specific geometry studied in the experiments and in this 
paper) the transition may become second order after going through a 
tricritical point. In general, though mean field theories are applicable 
in any dimension, the results are expected to be accurate only for higher
dimensions and long ranged potentials. The validity of the predictions of 
such theories for the system under consideration is, therefore, in doubt.  

A more recent theory~\cite{FNR} extends the dislocation unbinding mechanism
for two~-dimensional melting~\cite{KTHNY} for systems under external 
potentials. For a two~-dimensional triangular solid subjected to an external 
one~-dimensional modulating potential, the only dislocations involved are 
those which have their Burger's vectors parallel to the troughs of the 
potential. The system, therefore, maps onto an anisotropic, scalar Coulomb 
gas (or XY model)~\cite{FNR} in contrast to a {\em vector} Coulomb 
gas~\cite{KTHNY} for 
the pure $2~-d$ melting problem. Once bound dislocation pairs are integrated 
out, the melting temperature is obtained as a function of the renormalized 
or ``effective'' elastic constants which depend on external parameters like 
the strength of the potential, temperature and/or density. Though 
explicit calculations are possible only near the two extreme limits of
zero and infinite field intensities one can argue effectively that a 
re~-entrant melting transition is expected on general grounds  
quite independent of the detailed nature of the interaction potential 
for any two~-dimensional 
system subject to such external potentials. The actual extent of this region
could, of course, vary from system to system. In addition, these authors 
predict that the auto~-correlation function of the Fourier components of 
the density (the Debye~-Waller correlation function) decays algebraically in 
the solid phase with a universal exponent which depends only on the geometry 
and the magnitude of the reciprocal lattice vector. 

Computer simulation results in this field 
have so far been inconclusive. Early simulations~\cite{CKSS} involving 
colloidal particles
interacting via the Derjaguin, Landau, Verwey and Overbeek (DLVO) 
potential~\cite{col} found a large re~-entrant region
in apparent agreement with later experiments. On closer scrutiny, though,
quantitative agreement between simulation and experiments 
on the same system (but with slightly different parameters) appears to be 
poor~\cite{BBL}. Subsequent 
simulations~\cite{DK,DSK,DSK2} have questioned
the findings of the earlier computation and the calculated phase diagram 
does not show a significant re~-entrant liquid phase. 

\begin{figure}[hbtp]
\begin{picture}(0,80)
\put(0,0) {\psfig{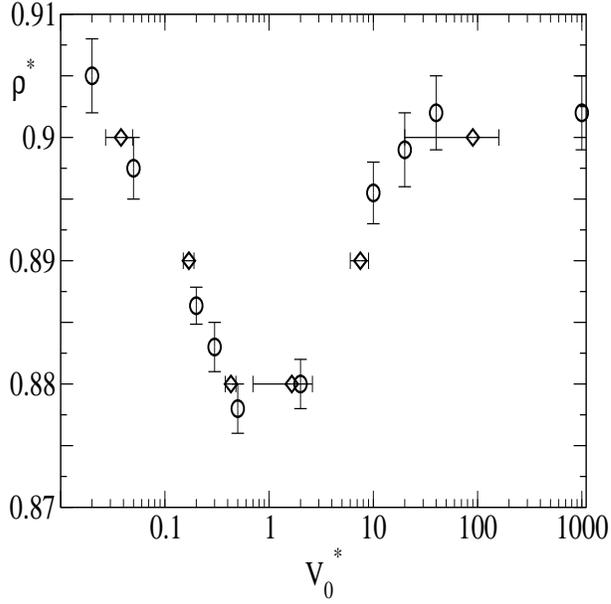}}
\end{picture}
\vskip .5 cm
\caption[]
{
~Phase diagram in the $\rho^\ast$ / $V_0^\ast$- plane.
Transition points for transitions from the solid to the modulated liquid
have been obtained by the order parameter cumulant intersection
method. In order to map the phase diagram we scanned in $\rho^\ast$ 
for every $V_0^\ast$, starting from the high density (solid) region.
The system size is $N = 1024$.
}
\label{phfig}
\end{figure}

Motivated, in part,
by this controversy, we have investigated the freezing/melting 
behaviour of an unrelated system subjected to similar modulating external 
potentials. In this paper we have computed the phase behaviour of a 
two dimensional hard disk system in an external potential.  
The pure hard disk system is rather well
studied~\cite{WB,henning,Jas,SNB} by now and the nature of the melting 
transition in the 
absence of external potentials reasonably well explored. Also, there 
exist colloidal systems with hard interactions~\cite{col}
so that, at least in principle, actual experiments using this system 
are possible. Finally, a hard disk simulation
is relatively cheap to implement and one can make detailed
studies of large systems without straining computational resources. 
The main outcome of our calculations, the phase diagram, is shown in
Fig.~\ref{phfig}.
We have shown results from  our simulation of a system of $N = 1024$ hard disks 
(diameter $\sigma$) of density 
$ 0.86 < \rho^* (= \rho \sigma^2) < 0.91$ and the amplitude of the external
potential $ 0 < V_0^* (= \beta V_0) < 1000$. Within our range of densities,
one has a clear signature of a re~-entrant liquid phase showing that this
phenomenon is indeed a general one as indicated in Ref.~\onlinecite{FNR}.

The rest of our paper is organized as follows. In Section II we specify the 
model and the simulation method including details of the finite size analysis
used. In Section III we present our results for the order parameter and its
cumulants with a discussion on finite size and hysteresis effects. We also 
present results for the specific heat, order parameter susceptiblity 
and correlation functions 
which further illustrates the nature of the phase transitions in this system.  
In Section IV we discuss our work in relation to the existing literature 
on this subject, summarize and conclude.

\section{Model and method}

\subsection{The Model}
We study a system of N hard disks of diameter $\sigma$ in a two dimensional
box of size $S_x \times S_y$ ($S_x/S_y = \sqrt{3}/2$) 
interacting with the pair potential $\phi(r_{ij})$
between particles $i$ and $j$ with distance $r_{ij}$,

\begin{equation}
\label{interaction}
%\phi(r_{ij})= \left\{ \begin{array}{r@{\quad:\quad}l}
\phi(r_{ij})= \left\{ \begin{array}{ll}
                 \infty & r_{ij} \leq \sigma \\
                 0      & r_{ij} > \sigma 
                 \end{array}
              \right.
\end{equation}  

\begin{figure}[hbtp]
\begin{picture}(0,80)
\put(0,-25) {\psfig{figure=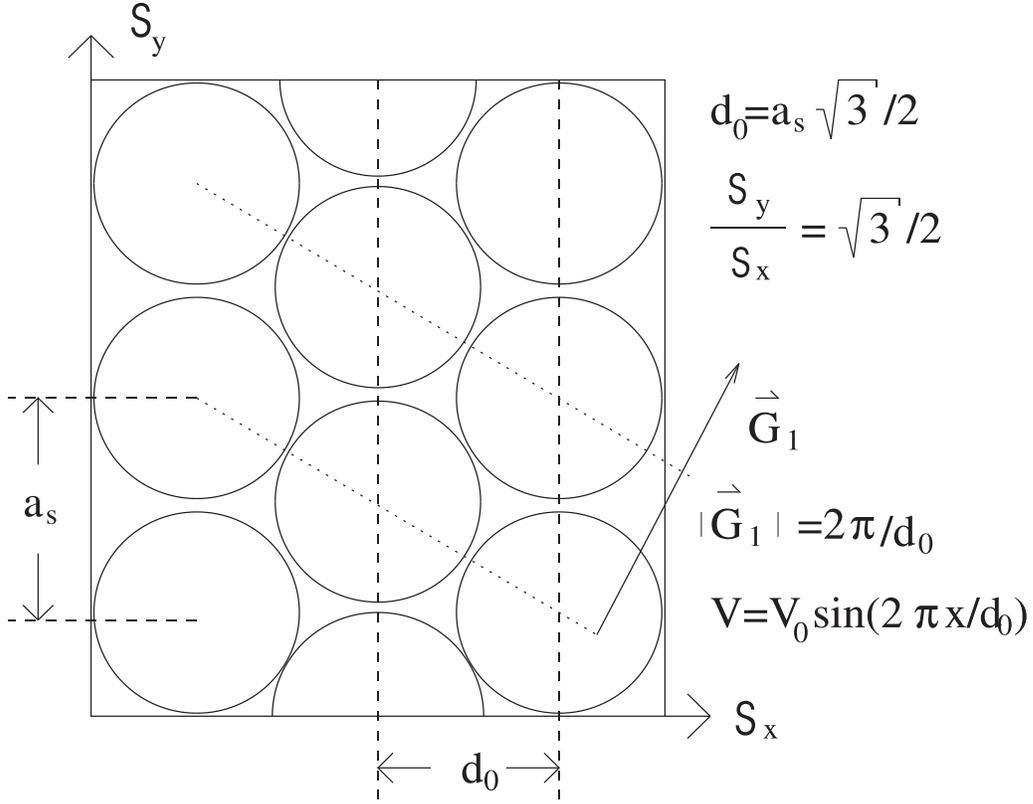,width=136mm,height=108mm}}
\end{picture}
\vskip 5 cm
\caption[]
{
~Schematic picture of the system geometry showing the direction
$\vec{G}_{1}$ along which crystalline order develops in the modulated liquid.
The four vectors obtained by rotating $\vec{G}_1$ anti~-clockwise by 
$60^\circ$ and/or reflecting about the origin are equivalent.
The parameters $d_0$ and $a_s$ are also shown. The size of the box is 
$S_x \times S_y$ and the modulating potential is $V$.
}
\label{gofig}
\end{figure}

In addition a particle with coordinates $(x,y)$ is exposed to an
external periodic potential of the form:
\begin{equation}
\label{field}
V(x,y) = V_0\, \sin\, (2 \pi\, x/d_0)
\end{equation}
The constant $d_0$ in Eq.(\ref{field}) is chosen such that, for a 
density $\rho= N/S_xS_y$, the modulation is commensurate to a triangular
lattice of hard disks with nearest neighbor distance
$a_s$: $d_0= a_s \sqrt{3}/2$
(see Fig.~\ref{gofig}).
The only parameters which define our system are the reduced density 
$\rho \sigma^2 = \rho^\ast$ and the reduced potential strength $V_0/k_B T = 
V_0^\ast$, 
where $k_B$ is the Boltzmann constant and $T$ is the temperature. 

\subsection{The Method}

\subsubsection{Numerical Details}
\begin{figure}[hbtp]
\begin{picture}(0,80)
\put(0,-54) {\psfig{figure=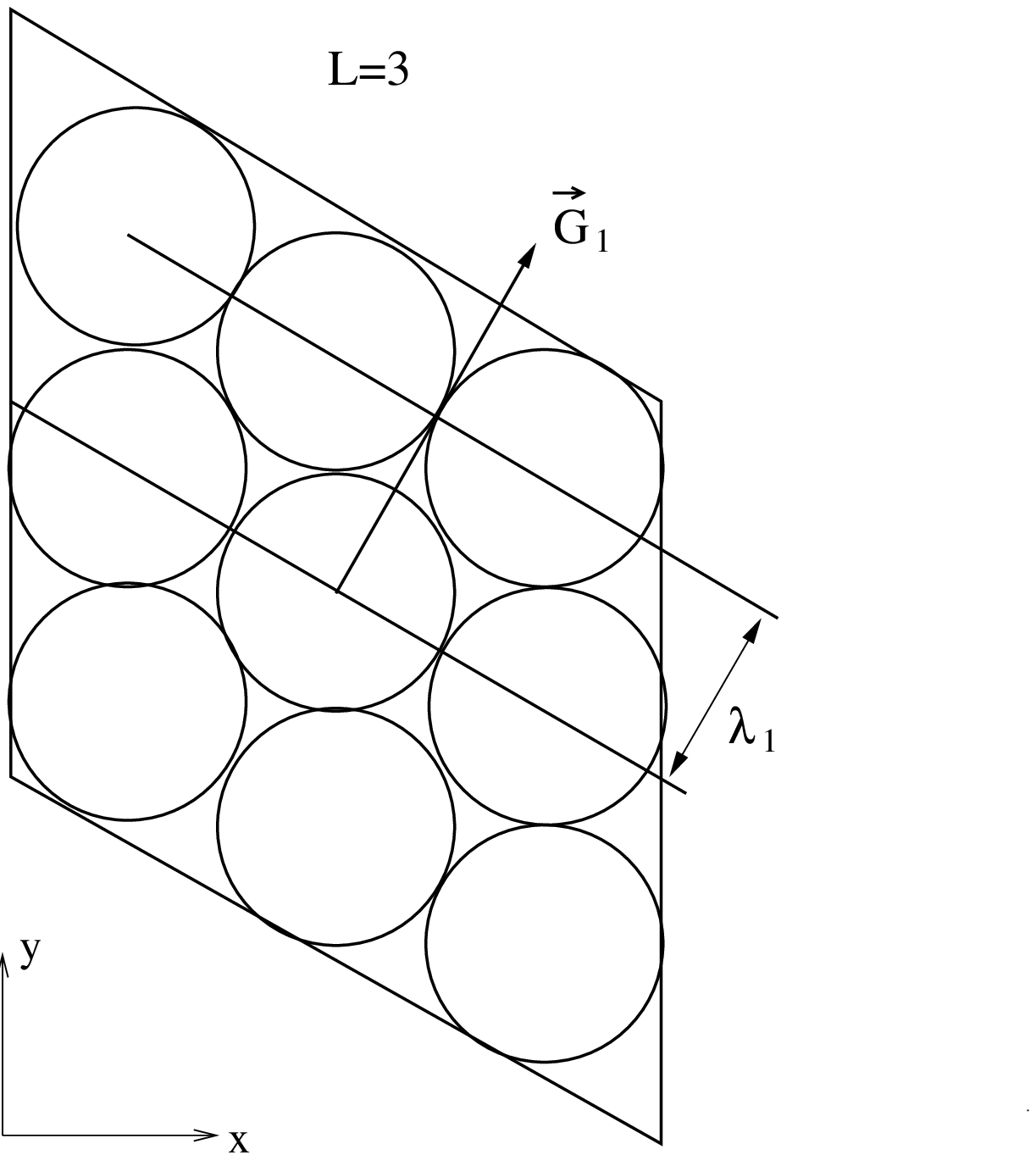,width=124mm,height=138mm}}
\end{picture}
\vskip 8 cm
\caption[]
{
~Schematic picture showing subboxes of size L (here $L=3$) 
used in the finite size scaling analysis (see text).
}
\label{g1fig}
\end{figure}

We perform NVT - Monte Carlo (MC) simulations~\cite{metro,LB}
for the system with 
interactions given by Eqs.~(\ref{interaction}) and (\ref{field})
for various values of  $V_0^\ast $ and $\rho^\ast$.
Averages $<\cdot>$ of observables have been obtained with the canonical 
measure. In order to obtain thermodynamic quantities for a range of system
sizes, we have analyzed various quantities within subsystems as shown in 
Fig.~\ref{g1fig}. We have used $<\cdot>_L$ to denote averages 
in subsystems. The subsystems are of size $L_x \times L_y$ where $L_x$
and $L_y$ are chosen as $L_y = L a_s$ and $L_x = L_y \sqrt{3}/2$
consistent with the geometry of the triangular lattice.

Most of the simulations described below have been done for $N=1024$ 
particles unless otherwise indicated. Phase transitions
have been studied in most cases by starting in the ordered solid
and reducing $\rho^\ast$ for fixed $V_0^\ast$. Runs where the density
$\rho^\ast$ is increased were also performed in a few cases. 

A typical simulation run with 4 $\times$ $10^7$ Monte Carlo steps (MCS)
(including 1.5 $\times 10^7$ MCS for relaxation)
took about 50 CPU hours on a PII/500 MHz PC.
At high values of $V_0^\ast$ in addition to ordinary (local) MC moves
we also used `through-moves', 
by which particle placements in neighboring troughs are tried.
Besides producing faster equilibration, including such moves ensures that 
the formation of dislocations for large $V_0^\ast$ 
and $\rho^\ast > \sqrt{3}/2$ ($d_0<\sigma$)
are not artificially hindered since particles can bypass each other
--- this is impossible with purely local MC moves.

\begin{figure}[hbtp]
\begin{picture}(0,80)
\put(0,0) {\psfig{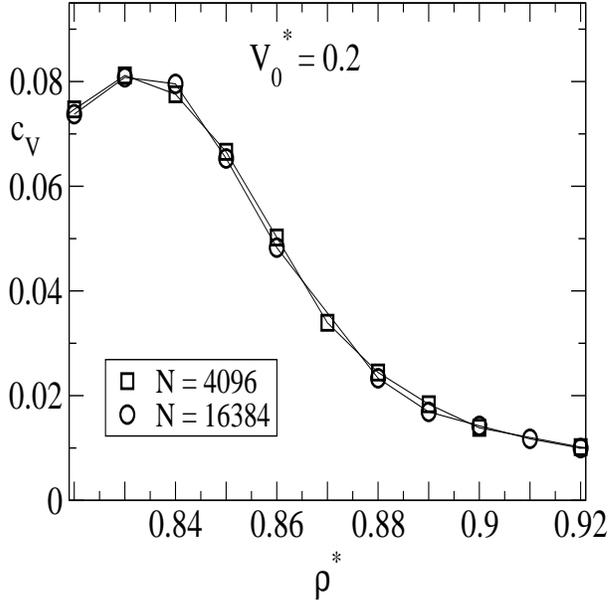}}
\end{picture}
\vskip .5 cm
\caption[]
{
~ Specific heat per particle versus density at constant $V_0^\ast=0.2$
and different system sizes ($N=4096$, $16384$).
}
\label{sphfig}
\end{figure}

To guarantee good equilibration and averaging, we simulated 
only systems up to N=1600 particles in the region of the phase boundary. 
Systems of N=4096 and N=16384 were used only once
in Fig.~\ref{sphfig}, where the interesting region is clearly 
in the liquid phase and equilibration is much easier.

\subsubsection{Order parameter}
The nature of the fluid-solid phase transition in two dimensions 
has been a topic of controversy throughout the last forty 
years~\cite{aldwain,KTHNY,ecstrand,henning,Jas,SNRB,SNB}.
%Long range orientational order can be found at high densities,
It is well known that true long range positional order is absent in 
the infinitely
large system due to low energy long wavelength excitations so that 
translational correlations decay algebraically. According to the 
dislocation unbinding mechanism~\cite{KTHNY,ecstrand} the two dimensional
solid (with quasi long ranged positional order) first melts into a 
``hexatic'' phase with no positional order but with quasi long ranged
orientational order signified by a non~-zero  bond orientational order parameter
$\psi_6 = \sum \exp(-i 6 \theta)$ where $\theta$ is the angle of a bond
and the sum is over all distinct bonds. A liquid,
with no bond orientational order either ($\psi_6 = 0$) is produced by a second
Kosterlitz~-Thouless (KT)~\cite{KTHNY} transition from the hexatic. 

In an external periodic field given by Eq.(\ref{field}), however, 
the bond orientational order parameter is non~-zero even in the fluid 
phase~\cite{FNR}.
This is because for  $V_{0}^{\ast} \neq 0$ we have now a ``modulated'' liquid, 
in which local hexagons consisting of the six nearest neighbors of a particle 
are automatically  oriented by the external field. Thus $<\psi_{6}>$
is non~-zero both in the (modulated) liquid and the crystalline phase and it
cannot be used to study phase transitions in this system.
The order parameters corresponding to a solid phase are the Fourier 
components of the (non~-uniform) density $\rho(\vec{r})$ calculated at the 
reciprocal lattice points $\lbrace \vec{G} \rbrace$. This (infinite) set 
of numbers are all zero (for $\vec{G} \neq 0$ ) in an uniform liquid 
phase and non~-zero in a solid. We restrict ourselves to the star consisting 
of the six smallest reciprocal lattice vectors of the two~-dimensional 
triangular lattice. In the modulated liquid phase that is relevant to our 
system, the Fourier components corresponding to two out of these six vectors,
viz. those in the direction perpendicular to the troughs of the external 
potential, are non~-zero~\cite{CKS}. The other four components of this set
consisting of those in the direction $\vec{G}_1$ (as defined for the ideal
crystal in Fig.~\ref{gofig}), and those 
equivalent to it by symmetry, are zero in the (modulated) liquid and 
non~-zero in the solid (if there is true long range order).
We therefore use the following order parameter: 
$$\psi_{G_1} = \left|\sum_{j=1}^N \exp (-i \vec{G}_1 \cdot \vec{r}_j)\right|$$
where $\vec{r}_j$ is the position vector of the $j^{th}$ particle. Note that 
though the order parameter  
$<\psi_{G_1}>$ decays to zero with increasing system size 
in the 2~-d solid  --- quasi long ranged order ---
this decay, being weak, does not hinder us from 
distinguishing, in a finite system, a modulated liquid from 
the solid phase with positional order in the $\vec{G}_1$ direction.

\subsubsection{Cumulants}
\label{cumul}
We have determined phase transition points by the
order parameter cumulant intersection method~\cite{KB}.
The fourth order cumulant $U_L$ of the order parameter distribution 
is given by:
\begin{equation}
U_L(V_0^\ast,\rho^\ast) = 1 - \frac{<\psi_{G_1}^4>_L}{3~<\psi_{G_1}^2>_L^2}
\end{equation}
In case of a continuous transition close to the transition point
the cumulant is only a function of the ratio of the system size 
$\approx L a_s$ and the correlation length $\xi$: $U_L(L a_s/\xi)$.
Since $\xi$ diverges at the critical point the cumulants for 
different system sizes intersect in one point: 
$U_{L_1}(0)= U_{L_2}(0) = U^\ast$.
Even for first order transitions these cumulants intersect~\cite{VRSB}  though 
the value $U^\ast$ of $U_L$ at the intersection is not
universal any more. The intersection point can, therefore,  be taken as 
the phase boundary regardless of the order of the transition.
This is useful since the order of the melting transition in $2~-d$ 
either in the absence~\cite{aldwain,KTHNY,ecstrand,henning,Jas,SNRB,SNB}
or with~\cite{CKS,CKSS,DK,DSK,DSK2,FNR} external potentials is not 
unequivocally settled.

In order to map the phase diagram we systematically vary the
system parameters $V_0^\ast$ and $\rho^\ast$
to detect order parameter cumulant intersection points which are then 
identified with the phase boundary.
It should be noted that though the order parameter (defined for
long range positional order) vanishes~\cite{KTHNY,ecstrand} 
with increasing system size in the crystalline phase, its
cumulants are well defined and can be used to determine phase boundaries.
For large $L$ the cumulants approach the value 2/3 in the solid phase
and 1/3 in the liquid~\cite{kumul} so that they are guaranteed to intersect~!
For very large $V_0^\ast$ we do not find an unique point of intersection 
for $U_L$, instead the cumulants for various values of $L$ collapse onto 
a single curve. In this case the 
onset of the collapse is taken as the ``intersection'' density.
It is curious to note that this behaviour is, in fact, typical of 
the anisotropic XY model~\cite{DL}.
In this case although the order parameter
cumulants have an intersection point, the value of the cumulant at the 
intersection differs for various anisotropies and drifts
towards a limiting value at zero anisotropy. The intersection ``point'' 
therefore changes to a ``line'' of intersections for different system 
sizes and for small anisotropies. In our system, for the large $V_0^\ast$
we see similar behaviour. 

\section{Results and Discussion}

\subsection{Order parameter and cumulants}
\label{opcu}

\begin{figure}[hbtp]
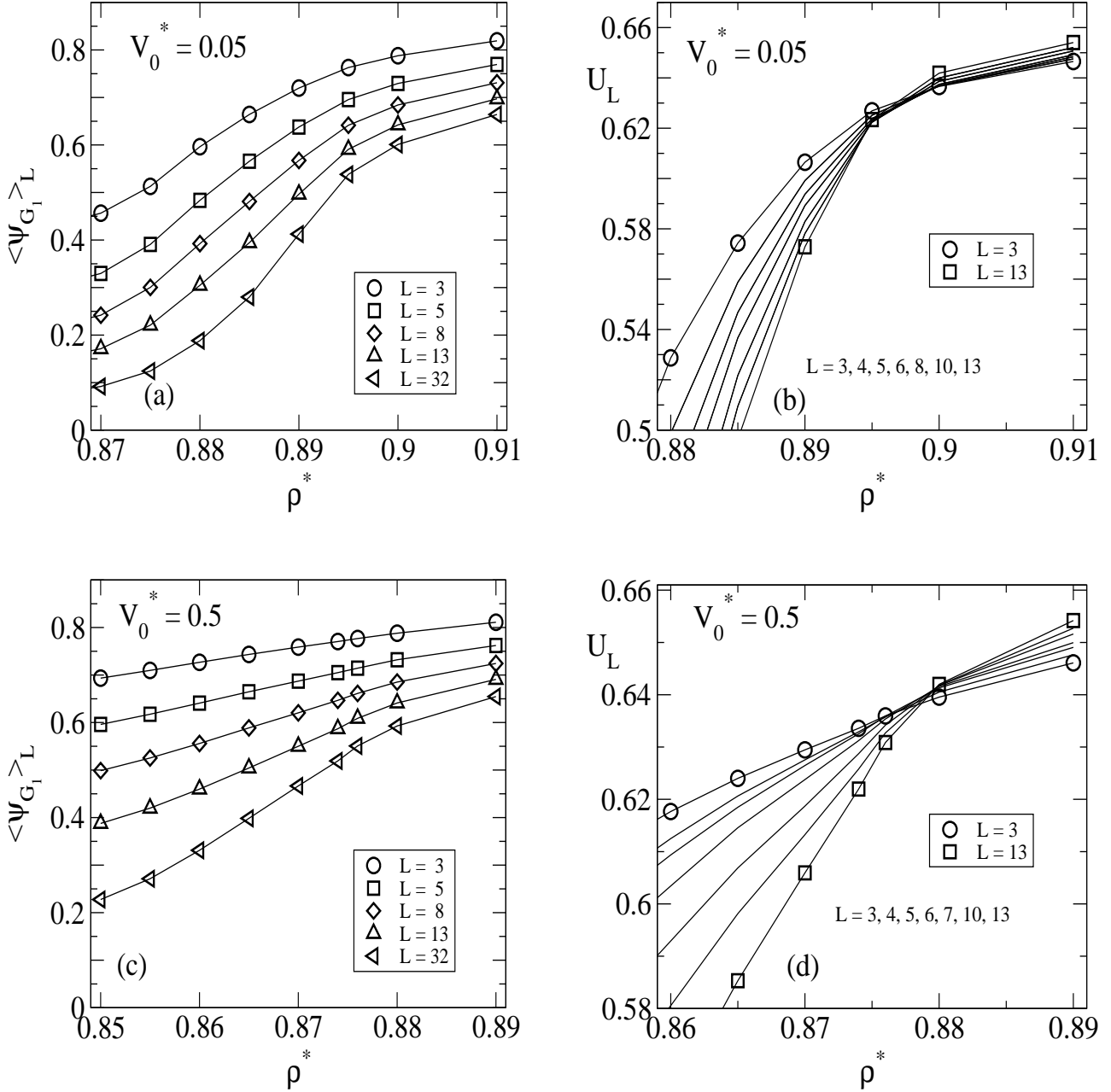

\begin{picture}(0,80)
\put(-10,0) {\psfig{figure=V0.05mittel_scale.eps,width=80mm,height=80mm}}
\put(80,0) {\psfig{figure=V0.05kumulanten.eps,width=80mm,height=80mm}}
\put(-10,-90) {\psfig{figure=V0.5mittel_scale.eps,width=80mm,height=80mm}}
\put(80,-90) {\psfig{figure=V0.5kumulanten.eps,width=80mm,height=80mm}}
\end{picture}
\vskip 12 cm
\caption[]
{
Order parameter $<\psi_{G_1}>_L$ ((a) and (c)) and order parameter cumulant
 $U_L$ ((b) and (d)) versus $\rho^\ast$
in subsystems of size $L$ for reduced potential amplitudes
$V_0^\ast = 0.05$ ((a) and (b)) and $V_0^\ast = 0.5$ ((c) and (d)), $N=1024$. 
Unless otherwise stated, lines connecting data points in this and the 
rest of the figures are for visual guidance.}
\label{opfig}
\end{figure}

In Fig.~\ref{opfig} we present data for the average order
parameter $<\psi_{G_1}>$ and its cumulants as functions of the density
for $V_0^\ast=0.05$ and $V_0^\ast = 0.5$ calculated within various 
subsystems.
In both cases $<\psi_{G_1}>_L$ and $U_L$ increase with $\rho^\ast$
with a sharpening of the structures for increasing $L$. 
As discussed above, we observe that for any density increasing subsystem size 
$L$ depresses  $<\psi_{G_1}>_L$. The cumulant $U_L$, on the other hand, 
approaches limiting values (2/3 for solid and 1/3 for liquid). 
The values of the cumulants are higher for
larger $L$ in the ordered (solid) phase and vice versa in the disordered 
(modulated liquid) phase thus resulting in an intersection point  --- the 
transition density.

\begin{figure}[hbtp]
\begin{picture}(0,80)
\put(0,0) {\psfig{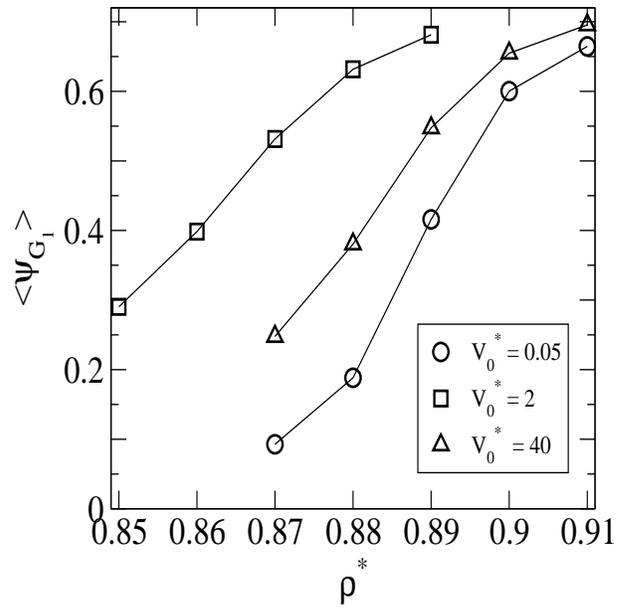}}
\end{picture}
\vskip .5 cm
\caption[]
{
~Average order parameter $<\psi_{G_1}>$ versus density for
$V_0^\ast = 0.05, 2,$ and $40$, the system size is $N=1024$.
}
\label{optwfig}
\end{figure}

In Fig.~\ref{optwfig} we compare the density dependence of 
the average order parameter $<\psi_{G_1}>$ (calculated over the entire 
system) for different
$V_0^\ast$ and for the same system size ($N=1024$).
With increasing $V_0^\ast$ the turning point in $<\psi_{G_1}>(\rho^\ast)$
is shifted to lower densities and then for even larger $V_0^\ast$ values
to higher densities. This indicates, already, that the system prefers having
smaller transition densities for intermediate values of $V_0^\ast$ compared
to smaller and higher $V_0^\ast$ values --- i.e. we have a re~-entrant 
transition. 

\begin{figure}[hbtp]
\begin{picture}(0,80)
\put(20,-30) {\psfig{figure=D0.89mittelwerte.eps,width=120mm,height=120mm}}
\put(20,-120) {\psfig{figure=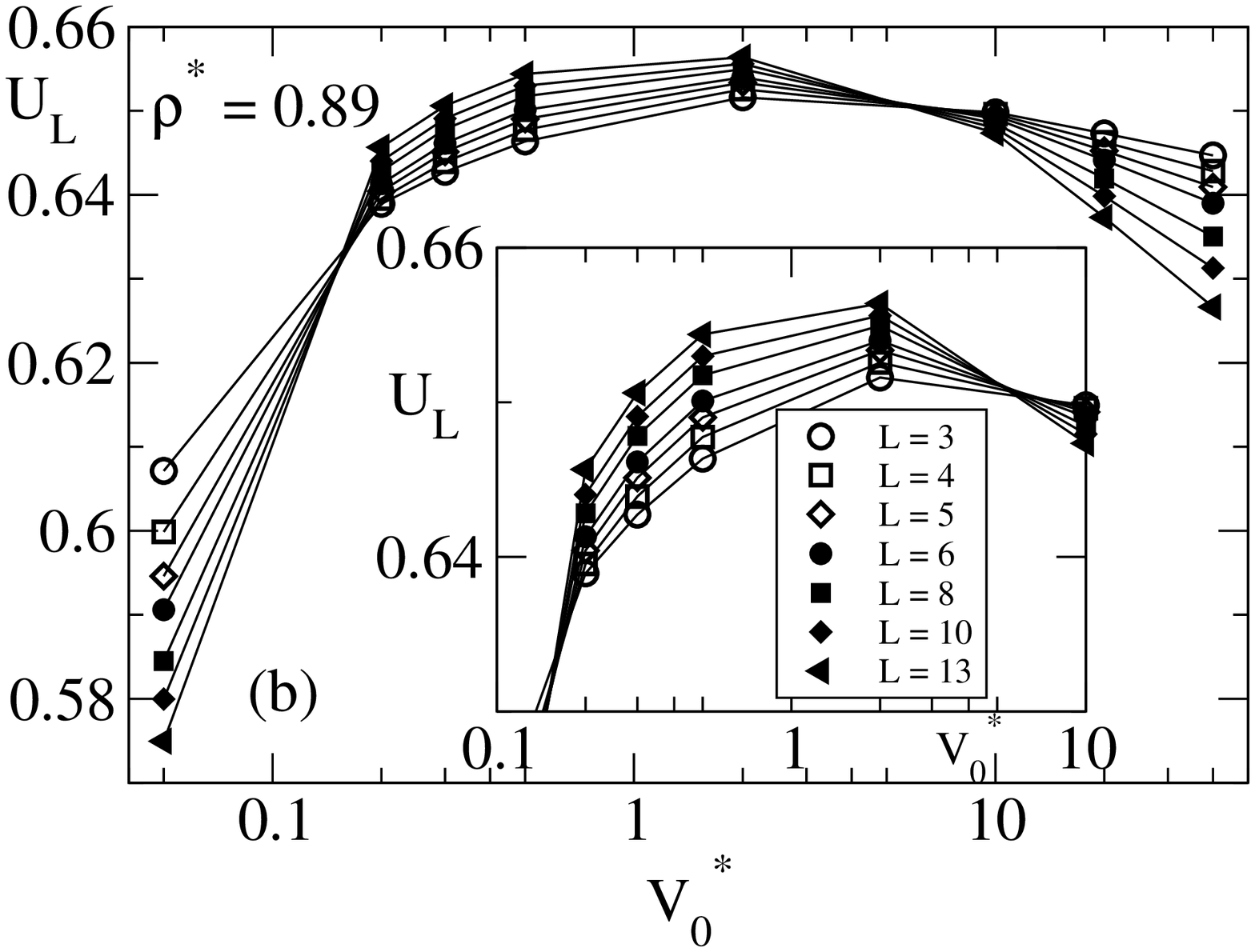,width=150mm,height=150mm}}
\end{picture}
\vskip 10.5 cm
\caption[]
{
~ Order parameter $<\psi_{G_1}>_L$ (a) and order parameter cumulant 
 $U_L$ (b) versus $V_0^\ast$ at a constant density $\rho^\ast = 0.89$
for different $L$, the system size is $N=1024$.
}
\label{op3fig}
\end{figure}

In Fig.~\ref{op3fig} we show a systematic study of $<\psi_{G_1}>_L$ and $U_L$ as 
a function of $V_0^\ast$ at the density $\rho^\ast=0.89$ for different
$L$- values. Maxima in the $<\psi_{G_1}>_L$- and $U_L$- curves are found
near $V_0^\ast= 2$. Again we note that the $<\psi_{G_1}>_L$- values
decrease with increasing $L$ (see Fig.~\ref{op3fig}(a)). The cumulants $U_L$, on the
other hand increases with $L$ for intermediate values of $V_0^\ast$ 
(the ordered, solid phase) and decreases with $L$ for either large 
or small $V_0^\ast$ (the disordered, liquid phase) resulting in intersection 
points indicating two consecutive phase transitions (see Fig.~\ref{op3fig}(b)). 

\begin{figure}[hbtp]
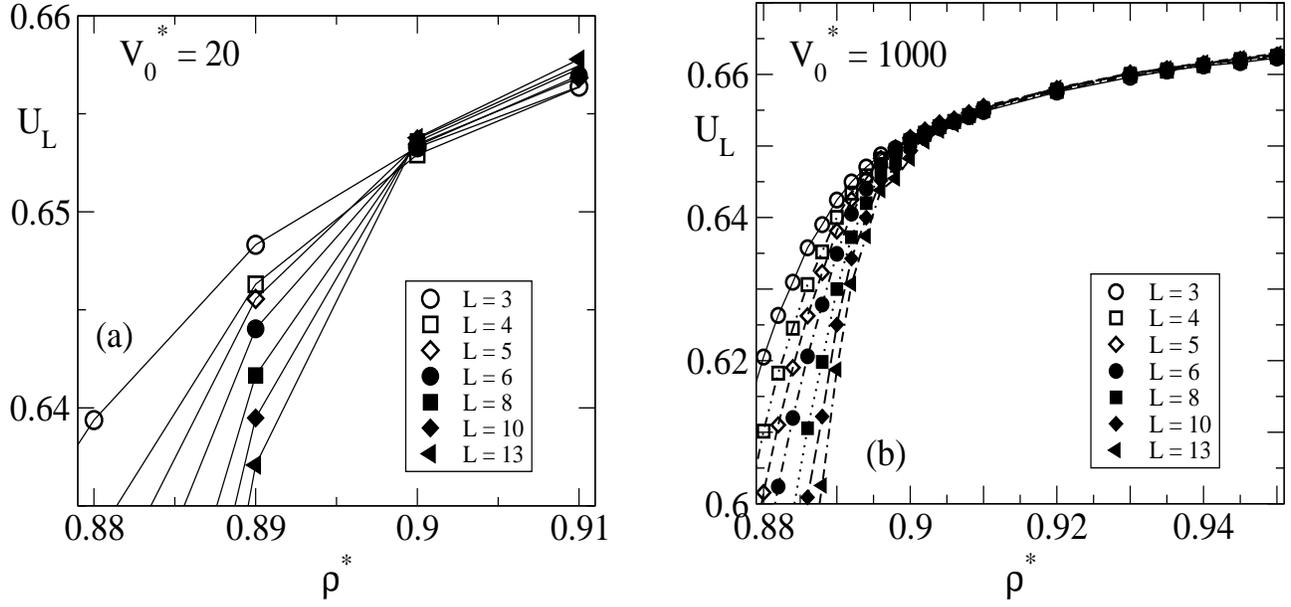

\begin{picture}(0,80)
\put(-10,0) {\psfig{figure=V20kumulanten.eps,width=80mm,height=80mm}}
\put(80,0) {\psfig{figure=V1000kumulanten.eps,width=80mm,height=80mm}}
\end{picture}
\vskip .5 cm
\caption[]
{
~
Order parameter cumulants $U_L$ versus $\rho^\ast$ at constant
$V_0^\ast = 20$ (a) and $V_0^\ast = 1000$ (b) for different $L$.
The system size is $N=1024$.
}
\label{op4fig}
\end{figure}

\begin{figure}[hbtp]
\begin{picture}(0,80)
\put(0,0) {\psfig{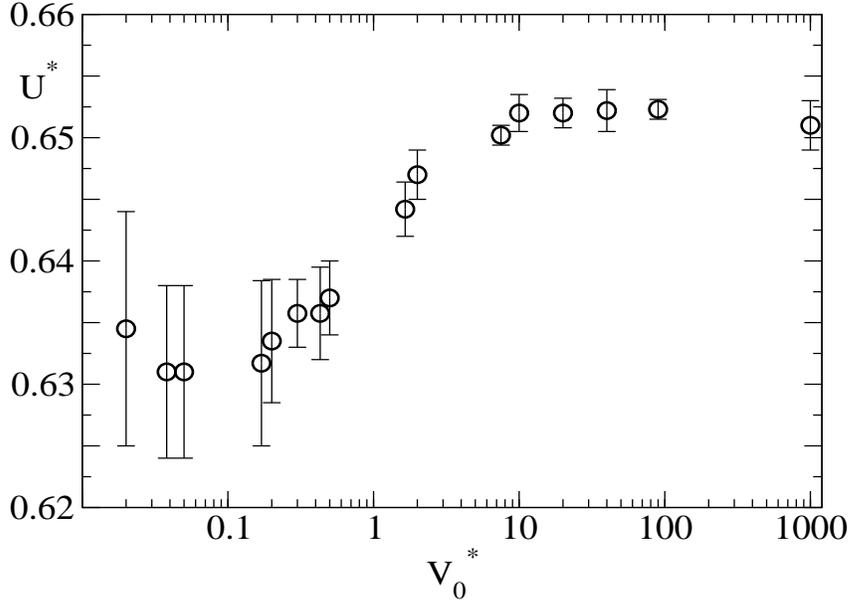}}
\end{picture}
\vskip .5 cm
\caption[]
{
~Values $U^\ast$ of the order parameter cumulants at the intersection
points versus $V_0^\ast$. The shown data at the largest four values of
$V_0^\ast$ are taken at the onset of the cumulants curves merging (see text).
The system size is $N=1024$.
}
\label{ufig}
\end{figure}

If $V_0^\ast$ is increased to $20$ the value of the cumulant 
at intersection $U^\ast$ is shifted upwards, see Fig.~\ref{op4fig}(a). For very high
$V_0^\ast$- values the cumulant curves for different $L$ merge 
on the high density side, see Fig.~\ref{op4fig}(b) (see discussion in Section \ref{cumul}).
In Fig.~\ref{ufig} the cumulant intersection values are shown as a function
of $V_0^\ast$, where for large $V_0^\ast$-values the value
at the onset of the merging is shown. We observe that $U^\ast$ is
not an universal number but, nevertheless, goes to a limiting value 
for large $V_0^\ast$~\cite{DL}. 

\begin{figure}[hbtp]
\begin{picture}(0,80)
\put(0,0) {\psfig{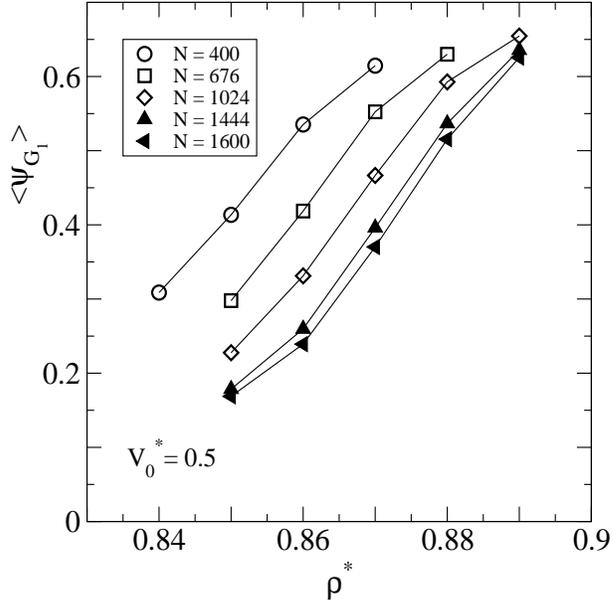}}
\end{picture}
\vskip .5 cm
\caption[]
{
~Order parameter $<\psi_{G_1}>$ versus density for different
system sizes ($N= 400,$ $676,$ $1024,$ $1444,$ $1600$) and $V_0^\ast=0.5$.
}
\label{op5fig}
\end{figure}

In Fig.~\ref{op5fig} we show $<\psi_{G_1}>$ as a function of the density with 
$V_0^\ast=0.5$ for different $N$-values. The general features
of $<\psi_{G_1}>$ as discussed above is retained though there is a shift
of the turning point to slightly higher densities with increasing $N$. 
The effect on the phase diagram is discussed in Section~\ref{phd}.

\subsection{Susceptibility, specific heat, finite size effects  and hysteresis}

In addition to $<\psi_{G_1}>_L$ and $U_L$ we have computed the
order parameter susceptibility $\chi_{G_1}$ and the specific heat
for different system- and subsystem-sizes.

\begin{figure}[hbtp]
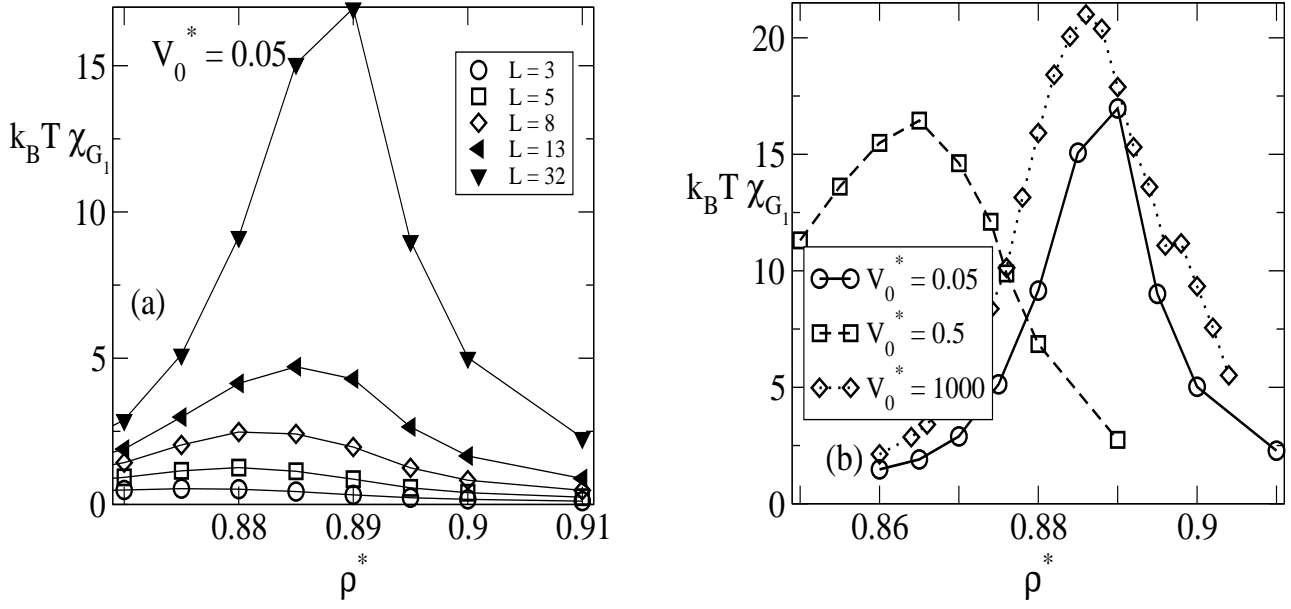

\begin{picture}(0,80)
\put(-10,0) {\psfig{figure=V0.05varianz_scale.eps,width=80mm,height=80mm}}
\put(80,0) {\psfig{figure=varianzen_b.eps,width=80mm,height=80mm}}
\end{picture}
\vskip .5 cm
\caption[]
{
~Order parameter susceptibilities versus density for:
(a) constant $V_0^\ast = 0.05$ and different $L$ values, $N = 1024$,
(b) full system size ($N=1024$) and different $V_0^\ast$- values.
}
\label{susfig}
\end{figure}

The order parameter susceptibility $\chi_{G_1}$ is defined as \cite{kumul}:
\begin{equation}
\label{suszep}
k_BT \chi_{G_1} = L^2 \left[ 
\left<\left(\psi_{G_1}\right)^2\right> - 
\left<\psi_{G_1}\right>^2 
\right]
\end{equation}
In Fig.~\ref{susfig}(a) we show $\chi_{G_1}$ as a function of $\rho^\ast$
at $V_0^\ast=0.05$ for different $L$- values. The increase
of $\chi_{G_1}$ with increasing $L$ signals the presence of
a phase transition in the density range where the transition
has been found by cumulant intersection techniques
($\rho^\ast_t \approx 0.896$). In Fig.~\ref{susfig}(b) $\chi_{G_1}$ is 
shown for the same system size ($N=1024$) and various $V_0^\ast$-values.
We note that the density of the $\chi_{G_1}$- maxima are smallest
for the intermediate value of $V_0^\ast$ which again show that for these 
$V_0^\ast$- values the transition density is lowest.
Compared to the cumulant intersection values, $\chi_{G_1}$ maxima are located at
slightly smaller densities (see also Section~\ref{phd}) which may be due to  
finite size effects, which often show the feature that 
phase transition points in finite systems
are shifted to slightly different values depending on the observable
under investigation. In particular one expects (and we get) a shift towards 
parameter values in the disordered region (here a liquid, i.e. low densities)
for the order parameter and the susceptibility as compared to the cumulant 
intersection parameters.

We have also calculated the specific heat $C_V$ (Fig.~\ref{sphfig}) as a function of 
the density for $V_0^\ast=0.2$ with $N=4096$ and $N=16384$. For a second 
order transition, the maximum of the specific heat scales with the system 
size as $C^{max}_V \sim L^{\alpha/\nu}$ where $\alpha$ and $\nu$ are 
critical exponents. For a first order transition, on the other hand 
$C^{max}_V \sim L^d$ where $d$ is 
the dimensionality ($= 2$ in our case). We, however, do not see any of 
this behaviour. In contrast, the specific heat is relatively featureless. 
Although it shows a peak, surprisingly, the height of this peak is almost 
insensitive to system size. This is a strong indication that the phase 
transition we observe is unconventional and is KT like~\cite{KTHNY,FNR}.
Further, as expected for such transitions, the maximum does not lie at the 
density where the 
cumulants intersect and it would be incorrect to identify specific heat 
maxima with the phase boundary (see discussion in Section \ref{concl}). 

\begin{figure}[hbtp]
\begin{picture}(0,80)
\put(0,0) {\psfig{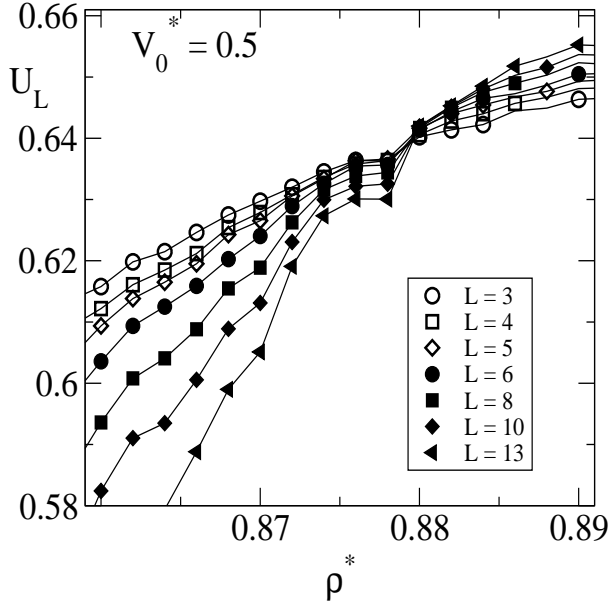}}
\end{picture}
\vskip .5 cm
\caption[]
{
~Order parameter cumulant $U_L$ versus density at constant $V_0^\ast = 0.5$
and different $L$, $N= 1024$. Values are obtained by successively compressing
the system from one density to the next higher density.
For a corresponding picture obtained by successively  expanding
the system see Fig.~\ref{opfig}(d).
}
\label{ucomfig}
\end{figure}

In order to study the effect of the path taken through the parameter space
on the location of the cumulant intersection densities, we
compared $U_L$ as a function of the density for $V_0^\ast=0.5$
as obtained from two runs. In Fig.~\ref{opfig}(d) we have already
presented the data for 
a run where the density was decreased systematically. In Fig.~\ref{ucomfig} we 
present data for a second run where the density was {\em increased}
instead.  We find negligibly small hysteresis effects on the cumulants as 
well as on the value of intersection density. This shows that the transition 
points are not affected much by the path through the parameter space. 

\subsection{The Phase Diagram}
\label{phd}
We have obtained the phase diagram of the system for $.86 < \rho^\ast < .91$
and $0 < V_0^\ast < 1000$.
For each density and $V_0^\ast$- value we computed 
cumulants $U_L$ for a range of subsystem sizes $L$ and located intersection 
points which we identify with the phase boundary.
The resulting locus of the transition points is shown in the phase diagram,
see Fig.~\ref{phfig}. At very small values of $V_0^\ast$
we find good agreement of our transition densities with the melting
densities ($\rho_m \approx 0.91$) known from literature~\cite{Jas}
on the pure hard disk solid ($V_0^\ast=0$).
The values of the transition density initially drop
and subsequently rise as $V_0^\ast$ increases.
The minimum transition density is found for $V_0^\ast \approx 1-2$.
These transition points separate a high density solid
from a low density modulated liquid. Thus, at a properly chosen density, 
we observe an initial  freezing transition followed by a re~-entrant melting 
at a higher $V_0^\ast$- value. Such an effect 
had been found earlier in experiments on colloidal systems
in an external laser field~\cite{bech,BWL,BBL}.

\begin{figure}[hbtp]
\begin{picture}(0,80)
\put(0,0) {\psfig{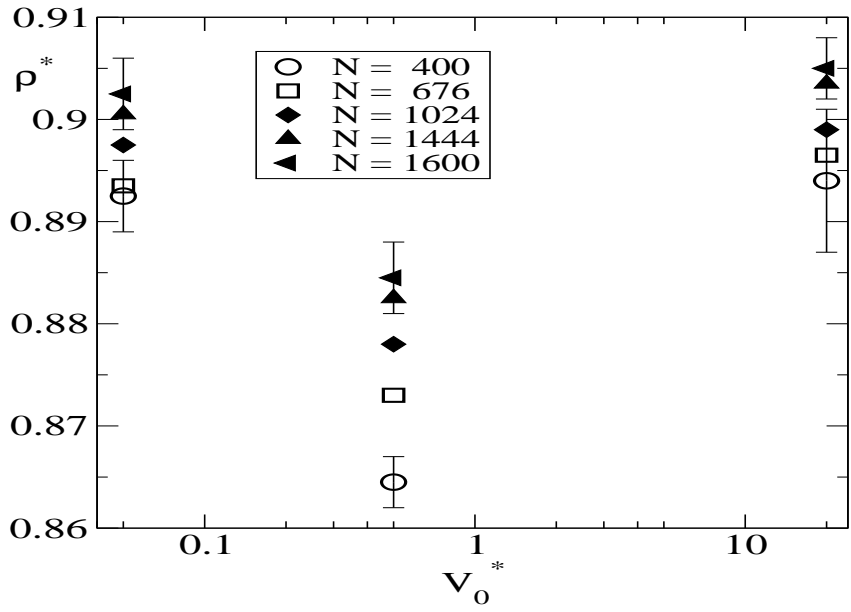}}
\end{picture}
\vskip .5 cm
\caption[]
{~Phase diagram in the $\rho^\ast$ / $V_0^\ast$- plane
for different system sizes ($N = 400,$ $676,$ $1024,$ $1600$).
}
\label{ph2fig}
\end{figure}

In order to quantify finite size effects on the phase diagram,
we have computed the transition points for different system sizes.
The resulting phase diagrams are shown in Fig.~\ref{ph2fig}. We note that due
to residual finite size effects with
increasing system size all transition points are slightly shifted to higher
densities, the structure of the phase diagram with a pronounced
minimum at intermediate values of $V_0^\ast$ is not affected by this shift.

\subsection{Correlation functions}

The Debye-Waller- correlation function
is defined as follows:

$$C_{\vec{G}_{1}}(\vec{R})=<e^{i\vec{G}_{1}(\vec{u}(\vec{R})-\vec{u}(0))}>$$

\noindent
where $\vec{R}$ points to the elementary cell of the ideal
lattice, and $\vec{u}(\vec{R})$ is the deviation of the actual particle
position from the ideal lattice:
$\vec{r}=\vec{R}+\vec{u}(\vec{R})$.      
In this case we have chosen the direction of $\vec{R}$ to lie along the
$y$ axis (i.e. along the troughs of the potential).

We have also computed the spatial correlation function $g(y)$ which is 
the pair correlation function in the $y$-direction. We compute it in the 
following way: for a particle i, $g(y)~dy~\propto$
number of particles j for which: 
$|y_{i}-y_{j}|\in[y, y+dy]$ and
$|x_{i}-x_{j}|<d_{0}/2$, 
normalized so that $g(y)\rightarrow 1$ as $y\rightarrow \infty$.

\begin{figure}[hbtp]
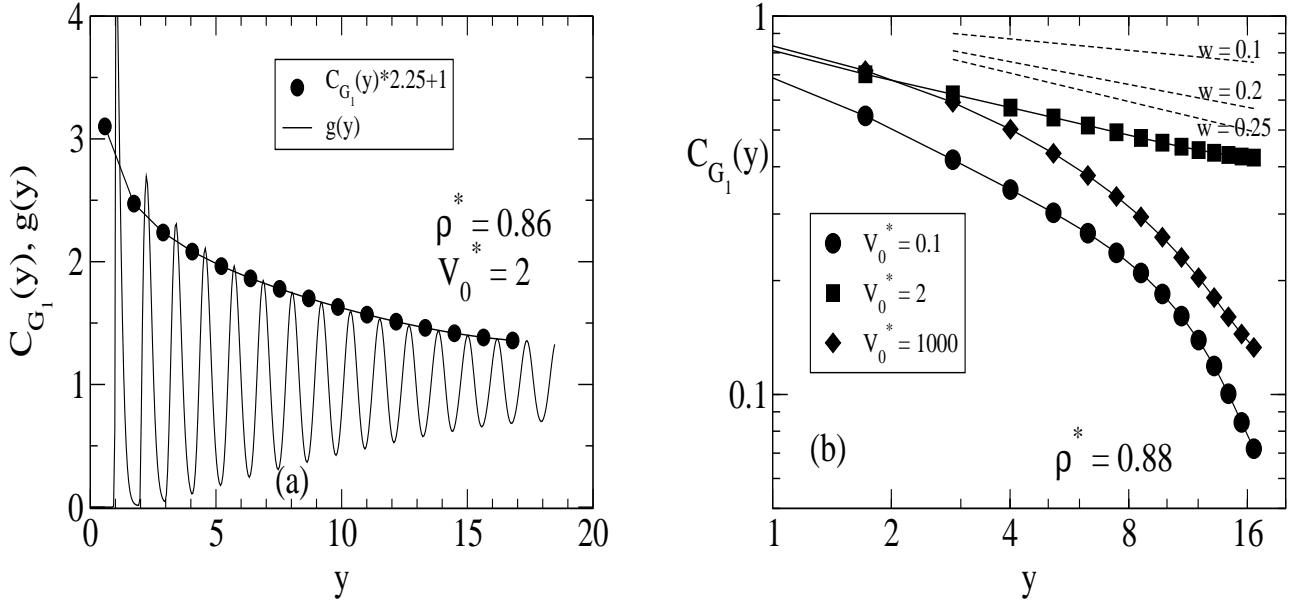

\begin{picture}(0,80)
\put(-10,0) {\psfig{figure=Debye_PaarKorrelation.eps,width=80mm,height=80mm}}
\put(80,0) {\psfig{figure=DebWallKorrelation.eps,width=80mm,height=80mm}}
\end{picture}
\vskip .5 cm
\caption[]
{
~Debye-Waller ($C_{G_1}(y)$) correlation- and pair distribution- ($g(y)$) 
functions as functions of $y$ parallel to the potential minima
for fixed $\rho^\ast=0.86$ and $V_0^\ast = 2$ (a)
and Debye-Waller correlation function versus $y$ for fixed
$\rho^\ast = 0.88$ and different $V_0^\ast = 0.1, 2, 1000$ (b).
Lines in the upper right corner of (b) show the functions $f(y) = y^{-{\rm w}}$ 
(${\rm w} = 0.1,$ $0.2,$ $0.25$) for comparison.
The system size is $N = 1024$.
}
\label{corfu}
\end{figure}

These correlation functions 
are plotted in Fig.~\ref{corfu} as functions of $y$. The Debye- Waller
correlation function $C_{G_1}(y)$ and the correlation function $g(y)$
along the potential valley are compared in Fig.~\ref{corfu}(a) at a density
just below the transition. We see that the decay of the maxima
of $g(y)$ as function of $y$ is similar to the decay in $C_{G_1}(y)$.
The decay of $C_{G_1}(y)$ is analyzed in more detail in Fig.~\ref{corfu}(b)
for parameter values in the liquid and in the solid phase.
In the liquid phase the decay is exponential while in the solid region
it is algebraic: $C_{G_1}(y) \sim y^{-\eta_{G_1}}$. Taking the data-points
in the crystal which are closest to the phase boundary for each 
$V_0^\ast$, we get $\eta_{G_1}$ in the range of $0.20 \dots 0.27$.
The exponent $\eta_{G_1}$ is predicted~\cite{FNR} to be universal and 
equal to $1/4$ for our geometry, so this value is consistent with 
our numerical results.

\subsection{Scaling behavior}

We next try to determine the order of the phase transitions encountered in this
system for various  values of $V_0^*$.  In order to investigate this 
issue we studied the scaling behavior
of the order parameter, susceptibility and the order parameter cumulant
near the phase boundary for a small ($0.5$) and a large ($1000$) $V_0$. 
From finite size scaling theory (for an overview
see Ref.~\cite{LB}) we expect these quantities to 
scale as~\cite{DL2} :

\begin{equation}
<\psi_{G_1}>_L L^b \sim f(L/\xi)
\label{scop}
\end{equation}
\begin{equation}
\chi_L k_BT L^{-c} \sim g(L/\xi)
\label{scchi}
\end{equation}
\begin{equation}
U_L \sim h(L/\xi)
\label{sccu}
\end{equation}
Here $b=\beta/\nu$, $c=\gamma/\nu$ (for critical scaling)
and $f$, $g$, $h$ are scaling functions.
The correlation length $\xi$ diverges as
$\xi \propto (1 - \rho/\rho_c)^{-\nu}$ for an ordinary critical point, while for
a KT- transition we have an essential singularity and
$\xi \propto \exp(a (1 - \rho/\rho_c)^{-\tilde{\nu}})$.

According to general arguments given in Ref.\cite{FNR}, we expect that 
for a finite lattice, the identification of the properties of our system 
with those of the anisotropic XY model should improve with increasing
$V_0^*$. Indeed, for large $V_0^*$, scaling according to the KT- theory seems
to be supported by our data. In Fig.~(\ref{sc1000}) we have plotted the left 
hand sides of Eqs.(\ref{sccu}) and (\ref{scchi}) versus $L/\xi$ for 
$V_0^*=1000$, where data points for $0.86 \leq \rho^* \leq 0.898$ have been 
considered and $\rho_c^* = 0.902$, obtained by cumulant intersection. In order 
not to introduce an unwarranted 
bias, we have separately considered (a) ordinary critical scaling and (b)
a KT scaling forms and adjusted the values of the parameters 
$b$, $c$ and $\nu$ till we obtained collapse of our data onto a 
single curve determined by a least square estimator.
Though good collapse of our data is observed 
both in (a) and (b), the numerical values for $\tilde{\nu}$, $\eta=2b$ and 
$c=2-\eta$ for KT scaling ($b \approx 0.138$, $c \approx 1.70$, $\tilde{\nu} 
\approx 0.44$) are close to the predicted values~\cite{FNR} 
($b = \eta /2 = 1/8$, $c = 1.75$, $\tilde{\nu} = 0.5$).   

The situation is less straight -forward for $V_0^* = 0.5$. 
The critical parameters were obtained 
in this case for densities $0.85 \leq \rho^* \leq 0.876$,
with  $\rho_c^\star = 0.878$.
In Fig.~\ref{sc05} (a) the data collapse looks slightly better than
in (b), such that relying on this data alone one may conclude that 
KT- scaling in this region of the phase diagram
seems less likely. It must be kept in mind though that for small values
of $V_0^*$ in a finite system, the analysis of the data would be complicated
by crossover effects. Strictly for $V_0 = 0$ we do not have a 
correspondence with the anisotropic XY model but rather with a {\em vector}
Coulomb gas\cite{KTHNY} with a different set of exponents. 
%Curiously the critical exponents in Fig~\ref{sc05} (a) ($V_0^* = 0.5$)  
%behavior are close to 2D-Ising values, though
%for $V_0^* = 1000$ $\nu$ is much larger than the 2D Ising value.
Our results for the numerical values of the parameters are summarized in
Table~I.

In summary, from the scaling analysis in Figs.~(\ref{sc05}) and (\ref{sc1000})
a KT- scenario at least for large $V_0^\ast$- values seems likely.
This is supported by the behavior of the cumulants as well (see
Sect.~\ref{opcu}).
A more precise classification of the phase transitions 
with the present data and system sizes is not easy.
This topic is left for future work, in particular 
we plan to compute the elastic properties of the system by a method
recently developed for the hard- disk system~\cite{SNB,SNRB} and to test the 
KT- predictions~\cite{FNR}.

\section{Summary and conclusion}
\label{concl}
In summary, we have calculated the phase diagram of a two dimensional system 
of hard disks in an external sinusoidal potential. We found  
freezing followed by re~-entrant melting transitions over a significant 
region of the phase diagram in tune with previous experiments on 
colloids~\cite{bech,BWL,BBL} and with the expectations of a recent dislocation
unbinding theory~\cite{FNR}. One of the main features of our calculation
is the method used to locate phase boundaries. In contrast to 
earlier simulations~\cite{CKSS,DK,DSK,DSK2} which used either the jump of 
the order parameter, or
specific heat maxima to locate the phase transition, we have used the 
more reliable cumulant intersection method. It must be noted that the 
specific heat in this system does not show a strong peak at the phase 
transition density so that its use 
may lead to confusing results. This, in our opinion, may be the reason for 
part of the controversy in this field. It is possible that earlier simulations
which used smaller systems and no systematic finite size analysis may have 
overlooked this feature of $C_V$ which becomes apparent only in computations 
involving large system sizes. We have shown that finite size scaling
of the order parameter cumulants as obtained from sub-system or
sub-block analysis, on the other hand, yields an accurate phase 
diagram.

What is the order of the phase transition? We know that~\cite{henning,Jas,SNB} 
for the pure hard disk system in two~-dimensions this question is 
quite difficult to answer and our present understanding~\cite{SNB} is that 
hard disks in two dimensions show a KTHNY transition. This transition, however,
lies very close (in an appropriately expanded parameter space) to a first order 
boundary so that crossover effects may be significant. The present system 
has an imposed external periodic potential which stabilizes 
the hexatic phase~\cite{FNR} and an (anisotropic) KT transition~\cite{FNR} 
is expected. 
Our results show several features which suggest that this is, perhaps,  what we 
have. Though we have discussed these observations in the rest of the paper, 
we list them below for clarity:

\begin{itemize}

\item
The behaviour of the value of the cumulants at intersection 
$U^\ast$ is similar to an earlier work~\cite{DL} on the anisotropic 
XY system which shows a KT transition. 

\item
The specific heat is relatively featureless and 
does not scale with system size in a fashion expected of a true first order or 
conventional continuous transition. 

\item
The decay of the correlation functions is similar to what is 
predicted~\cite{FNR} for an anisotropic scalar Coulomb gas.

\item
For large $V_0^*$- values the scaling of the order parameter, the 
susceptibility and the cumulant may be described by the KT- theory. 

\end{itemize} 

Of course, in order to resolve this issue unambiguously yet larger simulations
are required. Also, we need to compute elastic properties~\cite{SNRB,SNB} of
this system in order to compare directly with the results of Ref.~\cite{FNR}.
Work along these lines is in progress. 

Before we end we would like to point out that after completion of this work
and prior to the submission of this manuscript we received a 
preprint~\cite{PIN} where the same system as
ours has been studied using simulations. The phase diagram obtained by these 
authors is similar to ours (thus confirming and corroborating our results), 
though there exist some quantitative differences. These differences may be 
attributed to the absence of systematic finite size scaling in the latter work. 

\section{Acknowledgment}
We are grateful for many illuminating discussions with C. Bechinger 
and K. Binder.
One of us (S.S.) thanks the Alexander von Humboldt Foundation for
a Fellowship.
Support by the SFB~513 and granting of computer time
from the NIC and the HLRS is gratefully acknowledged.

%\end{multicols}

\begin{figure}[hbtp]
\begin{picture}(0,80)
\put(0,0) {\psfig{figure=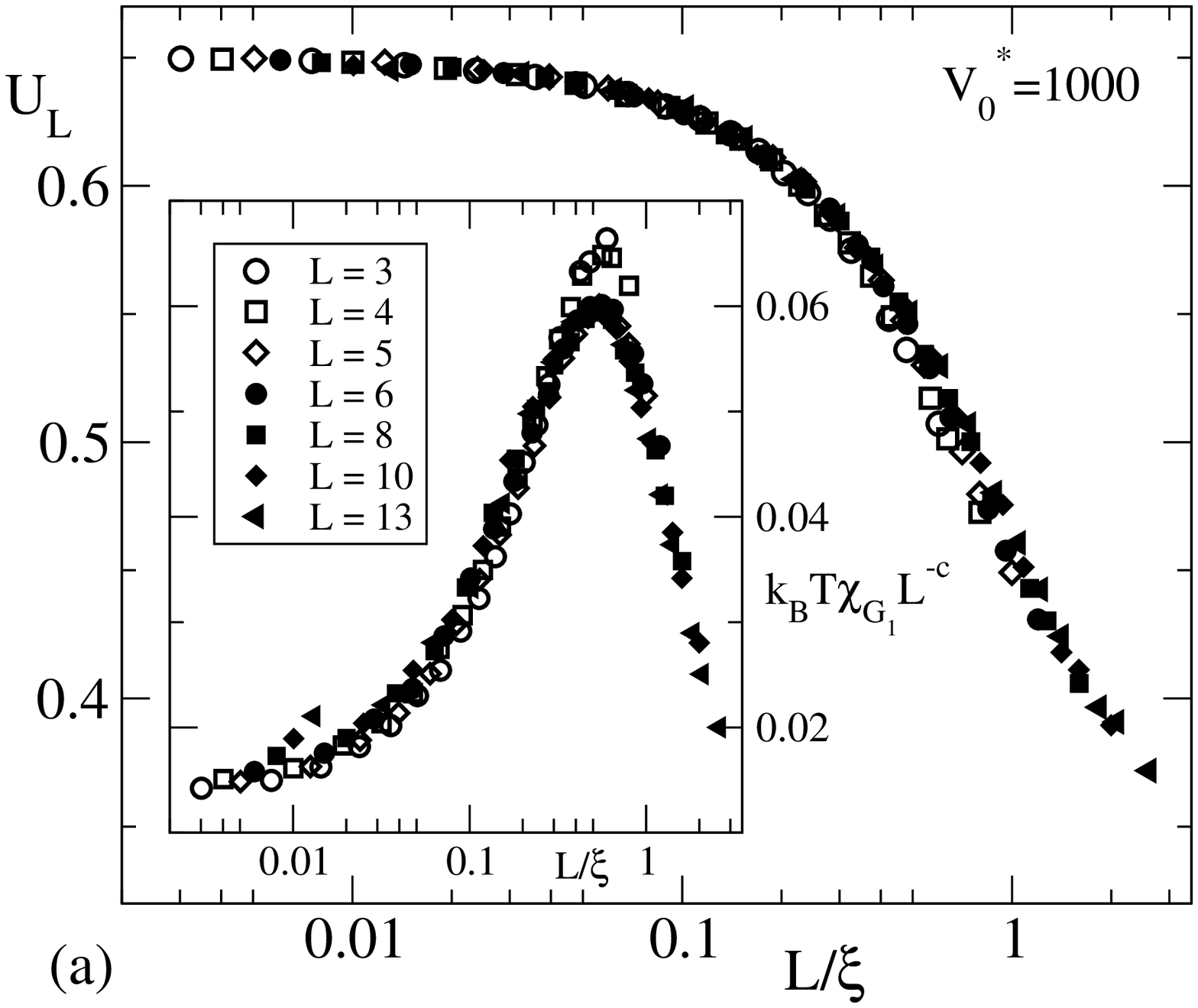,width=96mm,height=134mm}}
\put(0,-100) {\psfig{figure=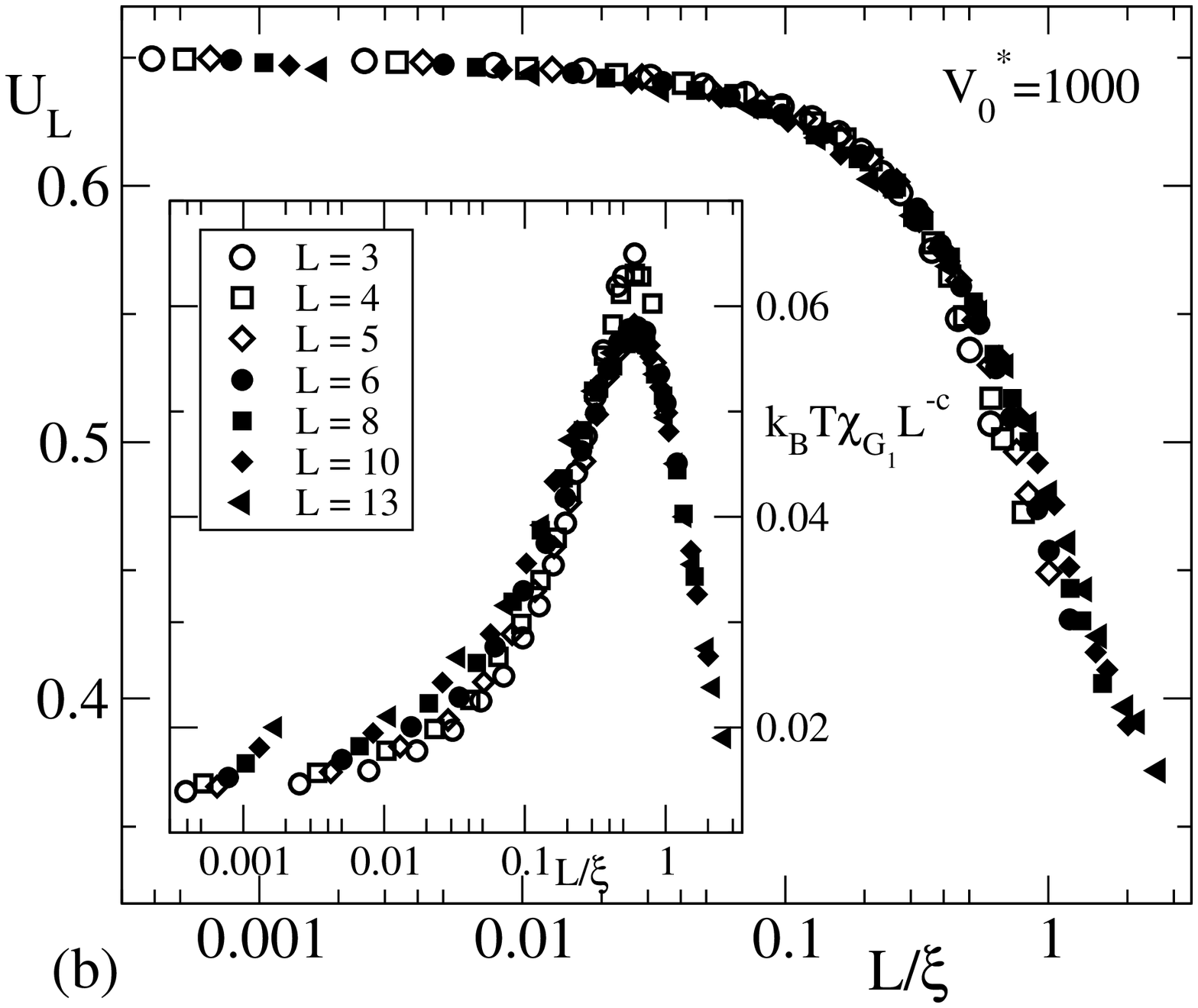,width=96mm,height=134mm}}
\end{picture}
\vskip 9.5 cm
\caption[]
{Scaling plots for the order parameter cumulant and the susceptibility (inset)
for $V_0^* = 1000$ assuming (a) critical and (b) KT- scaling.
The system size is $N = 1024$,
for $\xi$ we have used the expressions given after Eq.(\ref{sccu}).
}
\label{sc1000}
\end{figure}

\begin{figure}[hbtp]
\begin{picture}(0,80)
\put(0,0) {\psfig{figure=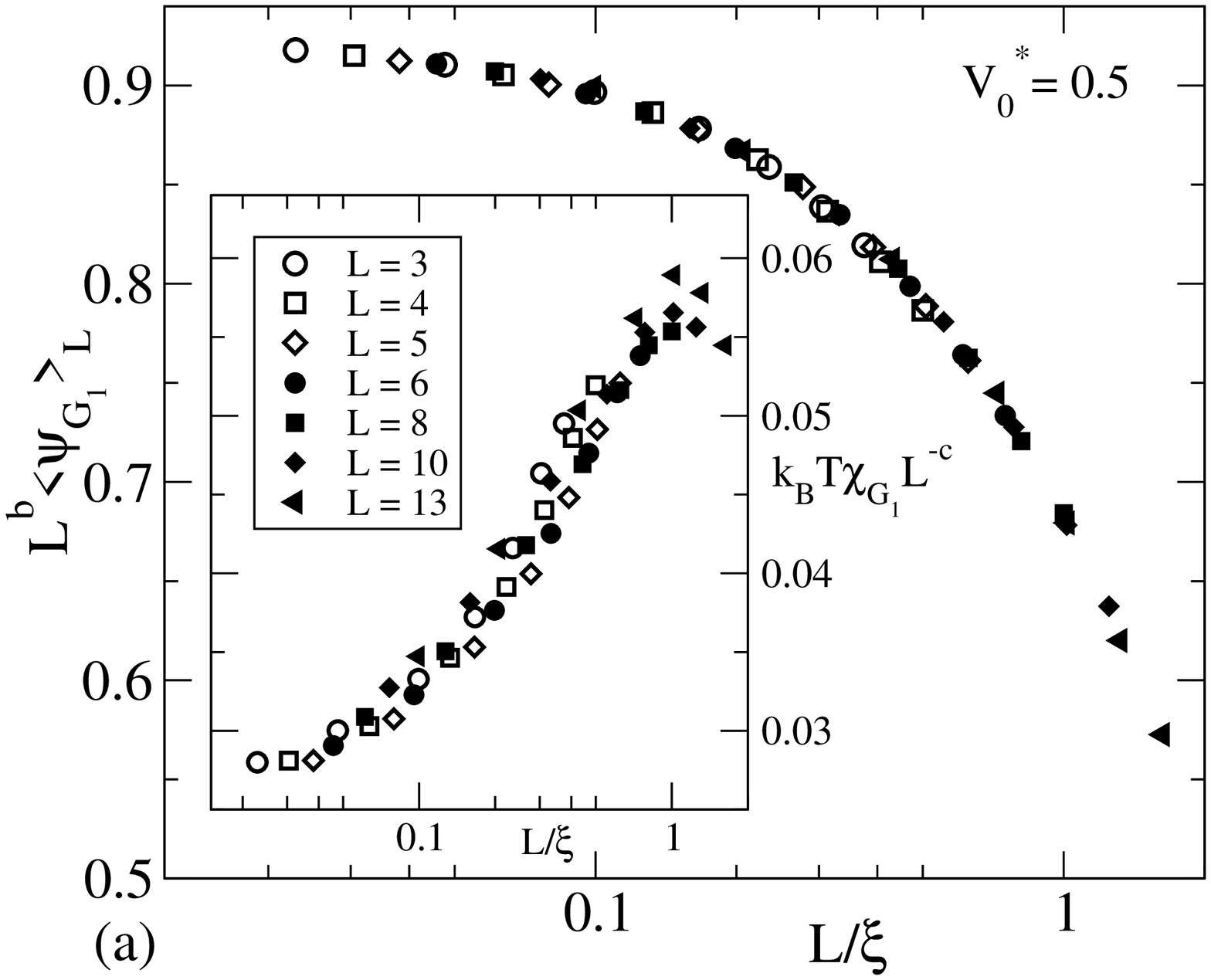,width=96mm,height=134mm}}
\put(0,-100) {\psfig{figure=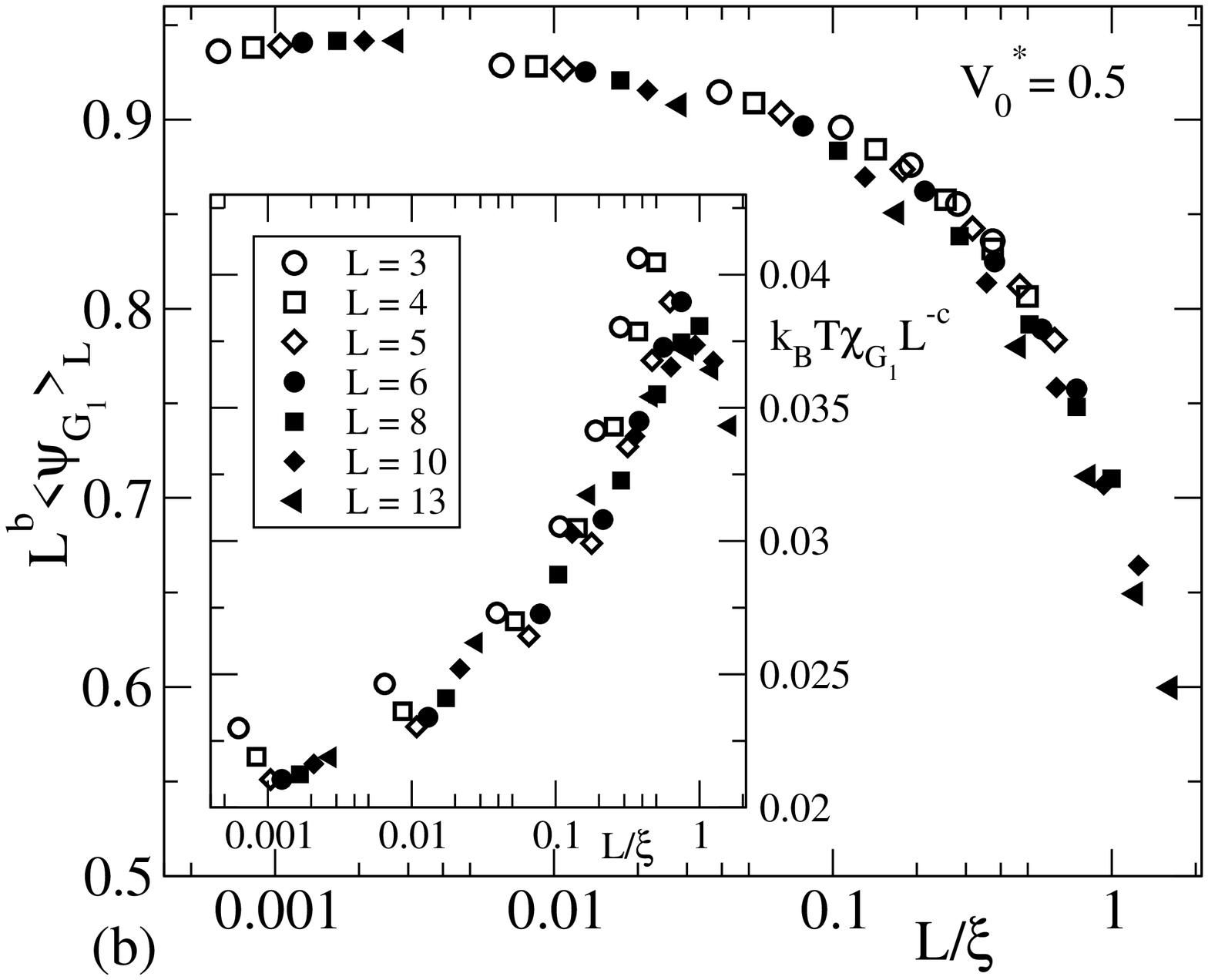,width=96mm,height=134mm}}
\end{picture}
\vskip 9.5 cm
\caption[]
{Scaling plots for the order parameter and the susceptibility (inset)
for $V_0^* = 0.5$ assuming (a) critical and (b) KT- scaling.
The system size is $N = 1024$,
for $\xi$ we have used the expressions given after Eq.(\ref{sccu}).
}
\label{sc05}
\end{figure}

\newpage
\begin{tabular} {||c|c|c|c||c|c|c|c||} \hline
%\begin{tabular} {| l l l l l l l |} \hline
$V_0^*$ & $b$ & $c$  & $\nu$ & $b$ & $c$  &  $\tilde{\nu}$ & $a$ \\ \hline
1000& 0.13(1) & 1.68(5) & 2.25(25)& 0.138(8) & 1.70(5) & 0.44(3) & 1.05(25) \\
0.5 & 0.152(5) & 1.65(6) & 1.06(13)& 0.170(12) & 1.83(4) & 0.38(10) & 1.0(2) \\ \hline
%0.05 & 0.135 & 1.51 & 2.1& 0.162 & 1.54 & 0.43 & 1.1(3) \\ \hline%
\end{tabular}
\vskip .5cm
\noindent
{\bf Table~I}~~ 
Parameters in the scaling plots (Figs.~(\ref{sc05}) and (\ref{sc1000}))
for $V_0^*=0.5$ and $V_0^*=1000$.
The first three parameter columns are for critical scaling, the last
four for KT- scaling.

\newpage

%\noindent
%{\bf \Large{Figure Captions}}
%\vskip 1cm
\newpage

\newpage

\newpage

\newpage

\newpage

\newpage

\newpage

\newpage

\newpage

\newpage

\newpage

\newpage

\newpage

\newpage

\newpage

\begin{references}
\bibitem{CAK} N.A. Clark, B.J. Ackerson, A. J. Hurd,
\prl {\bf 50}, 1459 (1983).

\bibitem{CAC} A. Chowdhury, B.J. Ackerson, N. A. Clark,
\prl {\bf 55}, 833 (1985).

\bibitem{LA} K. Loudiyi, B. J. Ackerson, Physica {\bf A 184}, 1 (1992); 
{\it ibid} 26 (1992).

\bibitem{bech} Q.~-H. Wei, C. Bechinger, D. Rudhardt and P. Leiderer, \prl
{\bf 81}, 2606 (1998).

\bibitem{BWL} C. Bechinger, Q.H. Wei, P. Leiderer, J. Phys.: Cond. Mat. 
{\bf 12}, A425 (2000).

\bibitem{BBL} C. Bechinger, M. Brunner, P. Leiderer, preprint.

\bibitem{maret} K. Zahn, R. Lenke and G. Maret, \prl {\bf 82}, 2721, (1999)

\bibitem{CKS} J. Chakrabarti, H.R. Krishnamurthy, A. K. Sood, \prl {\bf 73},
2923 (1994).

\bibitem{FNR} E. Frey, D.R. Nelson, L. Radzihovsky, \prl {\bf 83}, 2977 (1999).

\bibitem{CKSS} J. Chakrabarti, H.R. Krishnamurthy, A.K. Sood, S. Sengupta,
\prl {\bf 75}, 2232 (1995).

\bibitem{DK} C. Das, H.R. Krishnamurthy, \prb {\bf 58}, R5889 (1998).

\bibitem{DSK} C. Das, A.K. Sood, H.R. Krishnamurthy,
Physica {\bf A 270}, 237 (1999).

\bibitem{DSK2} C. Das, A.K. Sood, H.R. Krishnamurthy, preprint.

\bibitem{col} For an introduction to phase transitions in colloids see,
A. K. Sood in {\it Solid State Physics}, E. Ehrenfest and D. Turnbull Eds.
(Academic Press, New York, 1991); {\bf 45}, 1; P. N. Pusey in {\it Liquids,
Freezing and the Glass Transition}, J. P. Hansen and J. Zinn~-Justin Eds.
(North Holland, Amsterdam, 1991).

\bibitem{WB} K. W. Wojciechowski and A. C. Bra\'nka, Phys. Lett. {\bf 134A},
314 (1988).

\bibitem{henning} H. Weber, D. Marx and K. Binder, \prb {\bf 51}, 14636 (1995);
H. Weber and D. Marx, Europhys. Lett. {\bf 27}, 593 (1994).

\bibitem{Jas} A. Jaster, Phys. Rev. E {\bf 59}, 2594 (1999).

\bibitem{SNB} S. Sengupta, P. Nielaba, K. Binder,
\pre {\bf 61}, 6294 (2000).
 
\bibitem{metro} N. Metropolis, A. W. Rosenbluth, M. N. Rosenbluth, 
A. H. Teller, E. Teller, J. Chem. Phys. {\bf 21}, 1087 (1953).

\bibitem{LB} D.P. Landau, K. Binder, {\it A Guide to Monte Carlo
Simulations in Statistical Physics}, Cambridge University Press (2000).

\bibitem{aldwain} B. J. Alder and T. E. Wainwright, Phys. Rev. {\bf 127}, 359
(1962).

\bibitem{KTHNY} 
J. M. Kosterlitz, D. J. Thouless, J. Phys. {\bf C 6}, 1181 (1973); 
B.I. Halperin and D.R. Nelson, \prl {\bf 41}, 121 (1978);
D. R. Nelson and B. I. Halperin, Phys. Rev. B {\bf 19}, 
2457 (1979);
A.P. Young,
Phys. Rev. B {\bf 19}, 1855 (1979); K.J. Strandburg, Rev. Mod. Phys. {\bf 60},
161 (1988); H. Kleinert, {\em Gauge Fields in Condensed Matter}, Singapore,
World Scientific (1989).

\bibitem{ecstrand} K. J. Strandburg, \prb {\bf 34}, 3536 (1986).

\bibitem{SNRB} S. Sengupta, P. Nielaba, M. Rao and K. Binder,
\pre {\bf 61}, 1072 (2000).

\bibitem{KB} K. Binder, Z. Phys. {\bf B43}, 119 (1981);
K. Binder, \prl {\bf 47}, 693 (1981).

\bibitem{VRSB} K. Vollmayr, J.D. Reger, M. Scheucher, K. Binder,
Z. Phys. {\bf B 91}, 113 (1993).

\bibitem{DL} 
D.P. Landau, J. Magn. Magn. Mat. {\bf 31-34}, 1115 (1983))
 
\bibitem{DL2}
D.P. Landau, Phys. Rev. {\bf B27}, 5604 (1983).

\bibitem{PIN}
P. Chowdhury, A.K. Sood and H.R. Krishnamurthy, preprint.

\bibitem{kumul}
Since $\psi_{G_1}$ is the absolute value of a two-dimensional
quantity, we can assume the following probability distribution
in the (modulated) liquid and $L\gg\xi$ \cite{VRSB}:
$$w(\psi_{G_1})=\frac{\psi_{G_1}2L^{2}}{k_{B}T\tilde{\chi}}
exp\left(\frac{-\psi_{G_1}^{2}L^{2}}{k_{B}T\tilde{\chi}}\right)$$ 
which yields: $U_{L} = 1/3$ , 
$<\psi_{G_1}^{2}> = \frac{k_{B}T\tilde{\chi}}{L^{2}}$, and 
$\chi = \tilde{\chi}(1-\pi /4)$ for $\chi$ defined as in 
(\ref{suszep}).\\ 
Because $\psi_{6}$ is very similar in spirit, the same results
apply for $w(\psi_{6})$ in a isotropic liquid and $L\gg\xi$.


\end{references}
\end{document}